\documentclass[10pt, conference, compsocconf]{IEEEtran}

\usepackage{ifpdf}
\usepackage{cite}

\ifCLASSINFOpdf
  \usepackage[pdftex]{graphicx}
\else
  \usepackage[dvips]{graphicx}
\fi

\usepackage[cmex10]{amsmath}
\usepackage{algorithmic}
\usepackage{array}
\usepackage{mdwmath}
\usepackage{mdwtab}
\usepackage{eqparbox}
\usepackage[tight,footnotesize]{subfigure}
\usepackage{url}
\usepackage{braket}

\usepackage{tikz}
\usepackage{textcomp}
\usepackage{hyperref}
\usepackage{lipsum}
\newcommand\copyrighttext{%
  \footnotesize \textcopyright 2012 IEEE. Personal use of this material is permitted.
  Permission from IEEE must be obtained for all other uses, in any current or future
  media, including reprinting/republishing this material for advertising or promotional
  purposes, creating new collective works, for resale or redistribution to servers or
  lists, or reuse of any copyrighted component of this work in other works.
  DOI: \href{https://ieeexplore.ieee.org/document/9923816}{10.1109/MICRO56248.2022.00079}}
\newcommand\copyrightnotice{%
\begin{tikzpicture}[remember picture,overlay]
\node[anchor=south,yshift=10pt] at (current page.south) {\fbox{\parbox{\dimexpr\textwidth-\fboxsep-\fboxrule\relax}{\copyrighttext}}};
\end{tikzpicture}%
}

\begin{document}
\title{Q3DE: A fault-tolerant quantum computer architecture for multi-bit burst errors by cosmic rays}

\author{
\IEEEauthorblockN{Yasunari Suzuki\IEEEauthorrefmark{1}\IEEEauthorrefmark{2},
Takanori Sugiyama\IEEEauthorrefmark{3},
Tomochika Arai\IEEEauthorrefmark{1}\IEEEauthorrefmark{4}, 
Wang Liao\IEEEauthorrefmark{5},
Koji Inoue\IEEEauthorrefmark{6} and
Teruo Tanimoto\IEEEauthorrefmark{2}\IEEEauthorrefmark{6}
}
\IEEEauthorblockA{\IEEEauthorrefmark{1}
NTT Computer and Data Science Laboratories, NTT Corporation, Musashino, Tokyo 180-8585, Japan
}
\IEEEauthorblockA{\IEEEauthorrefmark{2}
JST, PRESTO, 4-1-8 Honcho, Kawaguchi, Saitama 332-0012, Japan
}
\IEEEauthorblockA{\IEEEauthorrefmark{3}
Research Center for Advanced Science and Technology (RCAST),\\ The University of Tokyo, Meguro-ku, Tokyo 153-8904, Japan
}
\IEEEauthorblockA{\IEEEauthorrefmark{4}
School of Science, The University of Tokyo, 7-3-1 Hongo, Bunkyo-ku, Tokyo 113-0033, Japan
}
\IEEEauthorblockA{\IEEEauthorrefmark{5}
Photon Science Center, The University of Tokyo, 2-11-16 Yayoi, Bunkyo-ku, Tokyo 113-8656, Japan
}
\IEEEauthorblockA{\IEEEauthorrefmark{6}
Faculty of Information Science and Electrical Engineering, Kyushu University, Fukuoka 819-0395, Japan
}
}

\maketitle
\copyrightnotice

\begin{abstract}
Demonstrating small error rates by integrating quantum error correction~(QEC) into an architecture of quantum computing is the next milestone towards scalable fault-tolerant quantum computing~(FTQC). 
Encoding logical qubits with superconducting qubits and surface codes is considered a promising candidate for FTQC architectures. 
In this paper, we propose an FTQC architecture, which we call \textit{Q3DE}, that enhances the tolerance to multi-bit burst errors~(MBBEs) by cosmic rays with moderate changes and overhead. There are three core components in Q3DE: in-situ anomaly DEtection, dynamic code DEformation, and optimized error DEcoding. In this architecture, MBBEs are detected only from syndrome values for error correction. The effect of MBBEs is immediately mitigated by dynamically increasing the encoding level of logical qubits and re-estimating probable recovery operation with the rollback of the decoding process. 
We investigate the performance and overhead of the Q3DE architecture with quantum-error simulators and demonstrate that Q3DE effectively reduces the period of MBBEs by 1000 times and halves the size of their region. Therefore, Q3DE significantly relaxes the requirement of qubit density and qubit chip size to realize FTQC. Our scheme is versatile for mitigating MBBEs, i.e., temporal variations of error properties, on a wide range of physical devices and FTQC architectures since it relies only on the standard features of topological stabilizer codes.
\end{abstract}


\IEEEpeerreviewmaketitle

\section{Introduction}
\label{sec:introduction}
Quantum error correction~(QEC)~\cite{nielsen2002quantum} is a vital technology to achieve fault-tolerant quantum computing~(FTQC) as quantum bits~(qubits) suffer from larger error rates than classical ones. The dominant errors of qubits stem from their lifetime and the infidelity of their controls, which appear as spatially and temporally independent errors.
With QEC techniques, we can reduce the effective error rate of a logical qubit to an arbitrarily small value by increasing the code distance. The number of required physical qubits for representing each logical qubit increases as physical error rates become large and the required logical error rate becomes small.
While the qubit interactions are typically restricted in nearest neighboring ones, surface codes~\cite{kitaev1997quantum,bravyi1998quantum} enable efficient encoding of logical qubits with the physical qubits fabricated on a two-dimensional~(2D) grid. 
Therefore, huge efforts have been paid to design a scalable FTQC based on surface codes~\cite{fu2018microarchitecture,holmes2020nisq+,fu2019control,fu2019eqasm,duckering2020virtualized,krinner2022realizing,huang2020fault,zhao2021realization}. 

One of the expected difficulties in the development of scalable FTQCs is temporally and locally {\it dependent} errors, i.e., {\it Multi-Bit Burst Errors (MBBEs)}.
In particular, an urgent example of MBBEs is those induced by cosmic-ray strikes~\cite{vepsalainen2020impact,wilen2021correlated,mcewen2021resolving,liu2022quasiparticle,martinis2021saving,iaia2022phonon,pan2022engineering,xu2022distributed} on superconducting qubits~\cite{devoret1985measurements,nakamura1997spectroscopy,arute2019quantum}.
The cosmic-ray-induced bit upset of the classical memory cells, e.g., static random-access memory (SRAM) cells, is widely known as soft error~\cite{abe2012multi,seifert2015soft}. Since superconducting qubits are more sensitive to the energy deposit by cosmic rays, cosmic-ray strikes result in drastic changes of error properties in a vast region. Another sticky issue is its temporal variation that does not appear in state-of-the-art classical SRAM cells. Once a cosmic ray hits the substrate of superconducting qubits, the qubits temporally remain in a state with a higher error rate. 
McEwen~\textit{et al.}~\cite{mcewen2021resolving} experimentally reported that the lifetime of a few qubits around the incident position decreases by one or two orders of magnitude, the effect lasts a few tens of milliseconds, and the frequency of cosmic-ray strikes is once per ten seconds in a 26-qubit region. As described later in Sec.\ref{sec:motivation}, an effective logical error rate increases by a factor of about 100 on average compared with the value estimated without considering burst errors. 
As explained in Sec.\,\ref{sec:applicability}, MBBEs are also expected in a variety of qubit devices and situations. Therefore, there is a strong demand for an MBBE tolerant FTQC design.

Since current QEC techniques stand on an assumption with time-independent error probabilities, they cannot handle such MBBEs efficiently. 
Straightforward ways to solve this issue are 1) to increase the default code distance to make logical error rates sufficiently small in the worst case, 2) to suppress cosmic rays' effect to a negligible level at a device-level technology, or 3) to mitigate MBBEs by statistical estimation after the execution.
However, the first method is extremely inefficient because it requires denser qubit integration or a larger chip size, which leads to more severe MBBE effects. Since the latency of several instructions is proportional to the code distance, such a naive solution significantly increases the required encoding levels and degrades the instruction throughput on encoded qubits. 
The second is still in investigation and uncertain in its feasibility. Since a large-scale qubit design requires mediating many trade-offs under constraints, an architecture-level technique that solves the transient errors while avoiding complicating the quantum processor design is necessary.
The last method, known as quantum error mitigation~\cite{temme2017error,endo2018practical,takagi2021fundamental}, can mitigate errors if they are a few times during a single-shot run, but the frequency of cosmic-ray strikes and the state-of-the-art resource estimation~\cite{gidney2021factor,babbush2018encoding} imply that these methods cannot treat the MBBEs.

To tackle this critical problem, we propose an MBBE tolerant FTQC architecture, called \textit{Q3DE}, including three key algorithms, in-situ anomaly DEtection, dynamic code DEformation, and optimized error DEcoding. The idea behind Q3DE is to detect MBBEs in terms of position and duration, and to apply space- and time-domain adaptive techniques to reduce the error rate. Our detection scheme does not require any additional action on qubits, i.e., they are detected only from the statistical changes of syndrome values. The dynamic code deformation attempts to temporally expand the code distance based on the duration of MBBE effects. The FTQC ISA is extended to support such function, i.e., a special instruction to modify the stabilizer map is dynamically inserted into an instruction queue. Our error decoding scheme makes it possible to roll the execution of an error-estimation task just back to before MBBE occurrence and re-execute to re-estimate one of the most probable errors by taking the effects of detected MBBEs into account. Since the two adaptive schemes are orthogonal, we can combine them to handle the spatial and temporal features of MBBEs. As far as we know, this is the first work that proposes an FTQC architecture to mitigate the effect of MBBEs by cosmic rays with moderate overhead. Our contributions are as follows.

\begin{itemize}
    \item We theoretically formulate the problems of existing FTQC architectures on MBBEs by cosmic rays. Statistical modeling of MBBEs is also presented. Then an MBBE tolerant FTQC architecture, Q3DE, is introduced with the detail of the algorithms and light-weight microarchitectural extensions of FTQC control units.
    \item The performance and overheads of Q3DE are evaluated with quantum-error simulators and logic synthesis with realistic parameters of cosmic rays. The results show that the dynamic code deformation suppresses the period exposed to an MBBE by about $10^3$ times, and the re-executed error decoding halves the size of the burst-error region.
    \item Our evaluation also shows that the combination of the adaptive deformation and decoding may achieve up to ten times reduction of required qubit count for a logical error rate below $10^{-10}$ and double the instruction throughput, compared with a naive solution that increases the default code distances.  
    \item As for the overheads, we have designed the introduced decoding unit by targeting an FPGA device. The results show that the hardware overhead caused by Q3DE is around 40\% in terms of LUT utilization and comparable hardware throughput, compared with a state-of-the-art baseline without MBBE consideration.
\end{itemize}

Although this paper focuses on the MBBEs by cosmic rays in superconducting qubits, the idea of Q3DE can be applied to a wide range of FTQC architectures to make the system tolerant to MBBEs, i.e., temporal variations of error properties. 
It is believed that the cosmic rays will incur similar errors in artificial qubits, such as color centers, quantum dots, and Majorana fermions~\cite{mcewen2021resolving}. Even for natural qubits (e.g., ions, neutral atoms, and photons), which would not suffer from cosmic rays, there are unavoidable temporal variations of error rates due to calibration drifts, qubit leakage, and so on.
Thus, MBBEs are essentially unavoidable in quantum computing, and quick detection and adaptive reaction to the temporal high-error-rate regions can be used as a versatile solution for ensuring the future scalability of FTQC.

\section{Background}
\label{sec:background}
\begin{figure*}
    \centering
    \includegraphics[width=1.0\textwidth]{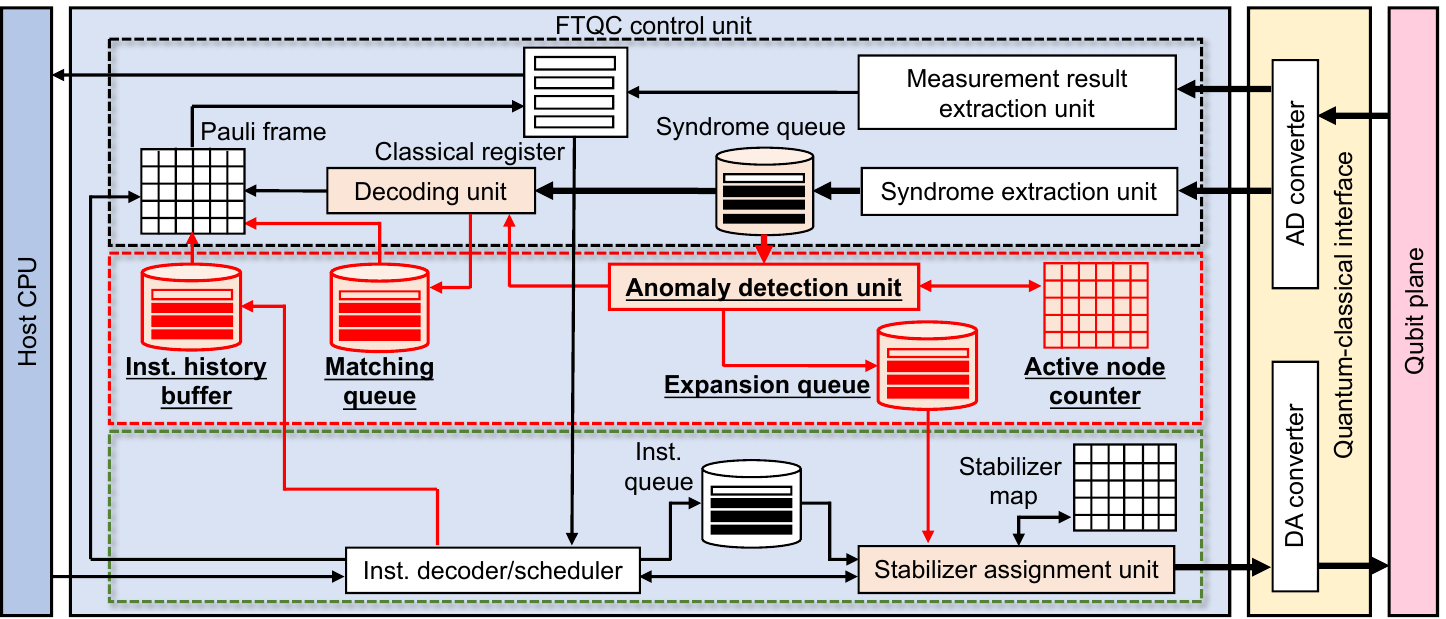}
    \caption{Design of our Q3DE architecture. Components in the blue region are classical computing units and those in the pink region are quantum computing units. The yellow region is their interface. Our proposal is to add components and interactions inside the red dotted square. We also modified the procedures of the units colored in orange.}
    \label{fig:architecture}
\end{figure*}

Our objective is to show that an extension of the FTQC architecture enables tolerance to MBBEs. Our design is shown in Fig.\,\ref{fig:architecture}. 
In this architecture, the components surrounded by the red dotted square are those added by our proposal. The whole architecture without this part constitutes a standard architecture of FTQC. We overview the elements of standard FTQC architectures in this section. For a more detailed introduction, see~\cite{fowler2012surface,fowler2018low}.
The notations used in this paper are summarized in Table\,\ref{tab:parameters}. 
\begin{table}
  \caption{Notations.}
  \label{tab:parameters}
  \centering
  \begin{tabular}{l|l}
  Notation & Description \\
  \hline
  $p$ & Physical error rate per code cycle \\
  $d$ & Code distance of logical qubits \\
  $p_{\rm L}$ & Logical error rate per code cycle \\
  $p_{\rm ano}$ & Physical error rate of anomalous qubits \\
  $d_{\rm ano}$ & Size of an anomalous region \\
  $p_{\rm L, ano}$ & Logical error rate with an anomalous region \\
  $c_{\rm win}$ & Cycle count of anomaly detection window \\
  $c_{\rm lat}$ & Cycle count of the latency of anomaly detection \\
  $d_{\rm exp}$ & Code distance of expanded logical qubits \\
  $p_{\rm L, opt}$ & Logical error rate with optimized error decoding\\
  \end{tabular}
\end{table}
We denote an addition modulo $2$ as $\oplus$. We denote a random variable with a hat ($\hat{\cdot}$) annotation.

\subsection{Quantum error correction}
\label{subsec:surface_code}
\begin{figure*}
    \centering
    \includegraphics[width=1.0\textwidth]{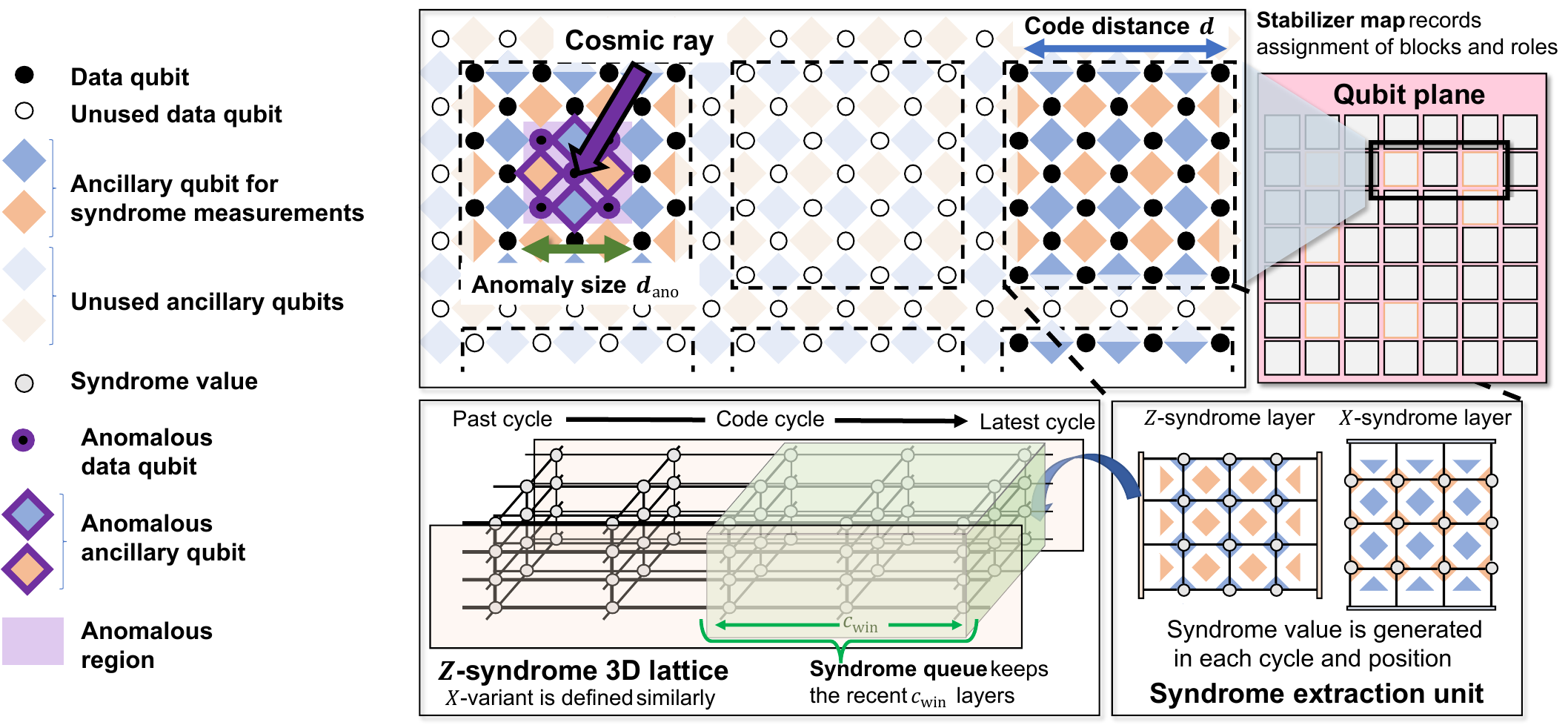}
    \caption{Schematic diagram of $d=4$ surface codes and the processing flow of syndrome values. Surface codes are allocated on the qubit plane. Z/X-syndrome values are extracted and stored in each buffer as 3D lattices in each surface code. Once a cosmic ray strikes the qubit plane, physical error rates of qubits around the hit position become high.}
    \label{fig:syndrome_flow}
\end{figure*}

We assume that physical qubits are fabricated on the \texttt{qubit plane} and allocated on a 2D grid and encoded with quantum error-correcting codes called surface codes~\cite{kitaev1997quantum,bravyi1998quantum,dennis2002topological,raussendorf2007fault,fowler2012towards}. Physical qubits in a square grid on the \texttt{qubit plane} are used for encoding a logical qubit, as shown in the upper right of Fig.\,\ref{fig:syndrome_flow}.
In the figure, black circles indicate data qubits storing the information of a logical qubit. Ancillary qubits allocated on red~(blue) squares are used to detect the parity of the number of Pauli-$X$~(-$Z$) errors acting on the data qubits at the corners, which is called a stabilizer measurement.\footnote{Pauli-$Y$ error is considered the combination of Pauli-$X$ and $Z$.}
We can detect any Pauli error acting on less than $d$ qubits in the allocated region, where $d$ is a code distance determined by the size of a region.
During the computation, we repetitively perform the stabilizer measurements to detect errors. The period of the repetition is called a code cycle and must be sufficiently shorter than the lifetime of qubits. The code cycle is typically assumed to be about $1~{\mu s}$ in the case of superconducting qubits~\cite{holmes2020nisq+,gidney2021factor,place2021new}. 

In each code cycle, from each ancillary qubit, the \texttt{syndrome extraction unit} extracts the parity of the number of errors on the monitoring data qubits, which is called a syndrome value. Then, we need to estimate a probable recovery Pauli operation\footnote{While errors are not necessarily Pauli, quantum states can be corrected by Pauli operations when a code distance is sufficiently large~\cite{suzuki2017efficient,bravyi2018correcting}.} from them. This process is called error decoding and is performed as shown in the bottom half of Fig.\,\ref{fig:syndrome_flow}. 
In each code cycle, we shape the obtained syndrome values as a layer, take an element-wise \textrm{XOR} for two consecutive layers, and stack it in a first-in-first-out~(FIFO) \texttt{syndrome queue}.
The data in the \texttt{syndrome queue} can be viewed as a three-dimensional~(3D) lattice, where each node corresponds to the parity of consecutive syndrome values and is called active when the parity is odd. Each edge corresponds to temporally and spatially local Pauli errors, and the occurrence of these errors is rephrased as the bit-flip of the nodes connected to the corresponding edge.
When the errors are assumed to be nearly local and uncorrelated, the estimation of a probable recovery Pauli operation is formulated as a minimum-weight perfect matching~(MWPM) problem, i.e., a task to match every active node to another active node or a boundary using a small number of edges~\cite{wang2003confinement,fowler2012towards,dennis2002topological}. 
We can suppress the probability of choosing a wrong recovery operation exponentially to the code distance as long as the physical error rate during each code cycle is smaller than a threshold value~\cite{fowler2012towards,holmes2020nisq+,ueno2021qecool,delfosse2017almost,das2020scalable,das2021lilliput}. This estimation is executed by the \texttt{decoding unit}. 
Since the active node may be matched to another active node observed in the future, the \texttt{syndrome queue} is expected to store no less than $d$ layers, and the latency of the \texttt{decoding unit} is at least $d$ code cycles. Once all the active nodes in the oldest layer are matched, we can remove the layer from the \texttt{syndrome queue}.
The estimated recovery Pauli operation is stored in the \texttt{Pauli frame}~\cite{fowler2012towards,riesebos2017pauli}. The information of the \texttt{Pauli frame} is used for correcting the binary outcome of logical measurements, as explained in the next section.

\subsection{Operations on encoded qubits}
\label{subsec:background_operations}
A universal set of quantum operations can be fault-tolerantly performed on logical qubits encoded with surface codes simply by changing the patterns of stabilizer measurements except for magic-state injection~\cite{fowler2018low,litinski2019game}. While there are a massive number of variants to execute logical operations, we focus on the succinct quantum instruction set shown in Table\,\ref{tab:instruction}, where \texttt{op\_expand} is an original instruction for mitigating the MBBEs.
\begin{table}
  \caption{A succinct instruction set for FTQC.}
  \label{tab:instruction}
  \centering
  \begin{tabular}{l|l}
  Name & Effect \\
  \hline
  \texttt{init\_zero} & Initialize a logical qubit in $\ket{0}$ states \\
  \texttt{init\_A} & Initialize a logical qubit in noisy $\ket{A}$ states \\ 
  \texttt{init\_Y} & Initialize a logical qubit in noisy $\ket{Y}$ states \\ 
  \texttt{op\_H} & Perform logical Hadamard gate \\
  \texttt{meas\_Z} & Measure a logical qubit in the Pauli-$Z$ basis \\
  \texttt{meas\_ZZ} & Measure two logical qubits in the Pauli-$ZZ$ basis \\
  \texttt{read} & \begin{minipage}{0.35\textwidth} Send an error-corrected measurement value\\from the classical register to the host CPU\end{minipage} \\
  \texttt{op\_expand} & Expand a code distance for mitigating MBBEs \\
  \end{tabular}
\end{table}
Note that we consider this set to concisely demonstrate the performance of our proposal. We expect our proposal to be applied to any instruction set of FTQC as long as it is built based on topological stabilizer codes. 

The instructions are pushed from the host CPU, cached in a FIFO \texttt{instruction queue}, and committed as soon as they are ready. All instructions except for \texttt{read} become ready if they commute with all the preceding and not-executed instructions and if there is an unused space for executing instructions on the qubit plane. The latter is arbitrated by the \texttt{stabilizer assignment unit} using the \texttt{stabilizer map}. Note that we can swap the order of error correction and these logical operations by updating the \texttt{Pauli frame}~\cite{aaronson2004improved,riesebos2017pauli}. 
When logical measurement instructions~(\texttt{meas\_*}) are executed, the raw output calculated by the \texttt{measurement result extraction unit} is sent to the \texttt{classical register}, which is marked as ``not-error-corrected''. When the \texttt{Pauli frame} catches up with the cycle of logical measurements, the entry of the \texttt{classical register} is corrected and marked as ``error-corrected''. A \texttt{read} instruction is a special instruction that requests to send the error-corrected outcome of logical measurements to the host CPU and does not request any action on the \texttt{qubit plane}.

Since almost all the logical instruction requires the unused blocks around the logical qubits or unused paths between target logical qubits, there is a trade-off relation between the density of used logical qubits and the throughput of instructions. Recent qubit-allocation strategies assume there are vacant blocks between each logical qubits~\cite{chamberland2022universal,beverland2022surface}. In this paper, we use the same qubit allocations to Ref.\cite{beverland2022surface}, i.e., we assume that blocks in even-indexed rows and columns on the qubit plane are used for logical operations.

\subsection{Impact of cosmic rays: SRAM vs. Superconducting Qubit}
\label{subsec:anomaly_qubit}
Recent studies have shown that cosmic rays temporally reduce the lifetime of superconducting qubits drastically, and the effect has become one of the main obstacles in the implementation of QEC~\cite{vepsalainen2020impact,wilen2021correlated,mcewen2021resolving,liu2022quasiparticle}. With small probabilities, cosmic rays hit the substrate of the \texttt{qubit plane} and scatter phonons. Since the energy of phonons is larger than a superconducting gap, the phonons generate a lot of quasi-particles in superconductors that can absorb the energy of qubits. This phenomenon shortens the lifetime of qubits around the incident position, as shown in the upper left of Fig.\,\ref{fig:syndrome_flow}.
Throughout this paper, we refer to qubits temporally affected by cosmic rays as {\it anomalous qubits}, the region of the incident as the {\it anomalous region}, and the size of an anomalous region as {\it anomaly size}.

The recent development of superconducting qubits has enabled experimental observation of the burst errors by cosmic rays. McEwen~\textit{et al.}~\cite{mcewen2021resolving} experimentally reported that the lifetime of anomalous qubits is reduced by one or two magnitudes of order, the anomaly size is about four in the current integration density, and the frequency of the burst errors is once in ten seconds on average in a 26-qubit region. The lifetime of anomalous qubits gradually recovers to the original lifetime, of which the decay constant is about $25~\mathrm{ms}$. 
This observation implies that cosmic rays incur widespread high-error-rate regions, which appears as an MBBE.

A similar phenomenon is known as a soft error in the existing classical memories. In the case of SRAM cells, the cosmic rays generate charged particles in the substrate. If the number of charged particles in the sensitive volume exceeds a critical charge, the classical bit will be flipped. The typical critical charge and sensitive volume depth of SRAM cells are evaluated as about $1~\mathrm{fC} \sim 22.5~\mathrm{keV}$ and $0.1~\mathrm{\mu m}$ order, respectively~\cite{abe2012multi,seifert2015soft}. 
As the density of SRAM cells increases with the scaling-down of technology, the proportion of multiple cell upsets~(i.e., simultaneous bit-flips in a single event) becomes larger~\cite{ibe2010impact}.
Compared to the classical counterpart, the superconducting gap of aluminum is about $0.36~\mathrm{meV}$~\cite{mcewen2021resolving}, and the diffusion radius of quasi-particles is about $6~\mathrm{mm}$~\cite{martinis2021saving}, both of which are several orders of magnitude worse than the classical ones. 
Thus, while the integration density of qubits is sparser than the classical ones, effects by cosmic rays would lose the reliability of FTQC, as in the case of high-performance computers.

\subsection{Related work}
\label{subsec:related_work}
Massive efforts have been made to quantify the MBBEs by cosmic rays in superconducting qubits~\cite{vepsalainen2020impact,wilen2021correlated,mcewen2021resolving,liu2022quasiparticle}. Throughout this paper, we refer to the parameters observed with Google's Sycamore chip~\cite{mcewen2021resolving} as a realistic assumption.
Various designs of superconducting qubits tolerant to cosmic rays have been discussed and proposed on the basis of this observation.
Martinis~\cite{martinis2021saving} investigated cosmic-ray-tolerant qubit designs with thick films of normal metal or low-gap superconductor. Pan~\textit{et al.}~\cite{pan2022engineering} and Iaia~\textit{et al.}~\cite{iaia2022phonon} experimentally reduced the effect of quasi-particles by using a metallic cover. 
Strikis~\textit{et al.}~\cite{strikis2021quantum} investigates the effect and mitigation of fabrication defects and mentioned that the QEC techniques for fabrication defects might be used for mitigating MBBEs. In the field of quantum networks, Xu~\textit{et al.}~\cite{xu2022distributed} proposed a scheme to mitigate MBBE effects by using distributed quantum error correction on multiple distinct chips and quantum communications.
Compared to these techniques, the advantage of our proposal is that it can detect MBBEs  non-destructively, can be implemented with modest changes to the classical post-processing, and can be completed in a single controller; i.e., it does not require any modification of the qubit design or quantum networks.
It should be noted that our technique does not conflict with any of the above works, and the adaptive QEC mechanism can be developed along with the physical qubit designs and combined with them.

\section{Motivation and overview of Q3DE}
\label{sec:motivation}
\subsection{The effect of multi-bit burst errors on FTQC}
\begin{figure}
    \centering
    \includegraphics[width=0.5\textwidth]{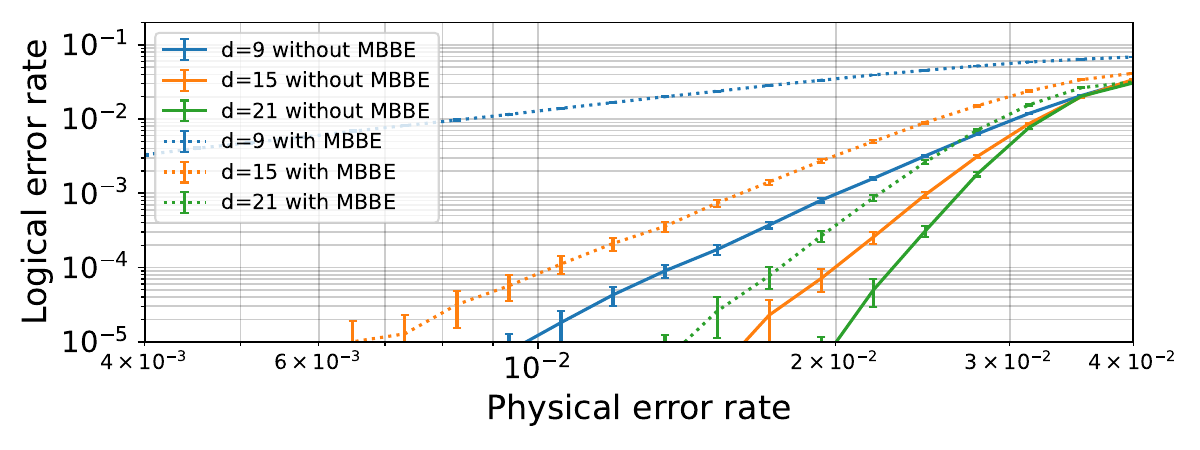}
    \caption{Logical error rates with and without an MBBE as a function of physical error rates. Dotted and solid lines indicate those with and without an MBBE, respectively.}
    \label{fig:logical_error_prob_motivation}
\end{figure}

We explain that a simple performance evaluation with realistic parameters indicates that the MBBEs significantly increase effective logical error rates.
We chose the parameters\footnote{Since several hundred physical qubits are used for a logical qubit in long-term applications, we multiplied the frequency $f_{\rm ano}$ by ten.} observed by McEwen~\textit{et al.}~\cite{mcewen2021resolving}: the frequency of cosmic rays is $f_{\rm ano}=1~\mathrm{Hz}$, and the effects of MBBEs last for $\tau_{\rm ano} = 25~\mathrm{ms}$.
Assuming that the multiple cosmic rays do not occur simultaneously on a chip, the effective logical error rate per cycle is 
\begin{align}
\label{eq:degradation_normal}
(1 - f_{\rm ano} \tau_{\rm ano}) p_{\rm L} +  f_{\rm ano} \tau_{\rm ano} p_{\rm L, ano}
\end{align}
on average, where $p_{\rm L, ano}$ and $p_{\rm L}$ are the logical error rate per cycle with and without an anomalous region.
In this equation, the second term is the contribution of MBBEs, and the effective increase ratio by cosmic rays is represented by $f_{\rm ano} \tau_{\rm ano} p_{\rm L, ano}/p_{\rm L}$.
We numerically evaluated $p_{\rm L}$ and $p_{\rm L, ano}$ for several code distances with $d_{\rm ano}=4$ and $p_{\rm ano}=0.5$. For the detailed simulation settings and the definition of parameters, see Sec.\,\ref{subsec:evaluation_simulation_settings}. The numerical results are shown in Fig.\,\ref{fig:logical_error_prob_motivation}. 
We can see that the occurrence of an MBBE significantly increases the logical error rates. The increase of error rates becomes significant as the physical error rate becomes lower.

Another important observation is that a single MBBE event does not change the threshold value of surface codes even though it incurs significant performance degradation. The physical error rate at which plots of sufficiently large code distances cross is called a threshold value, which is a popular measure to evaluate the error-correction performance~\cite{holmes2020nisq+,ueno2021qecool}. However, we observed that the threshold value is almost completely independent of the occurrence of an MBBE. This result implies that we need to investigate the MBBE effect on surface codes and propose reasonable methods to evaluate the performance degradation. This point is analyzed with the first-order approximation in Sec.\,\ref{subsec:decoder_analysis}.

A simple solution to MBBEs is to increase the default code distance to a sufficiently larger value than the anomaly size. However, this solution requires more physical qubits as an overhead, which leads to a decrease in the number of available logical qubits. Furthermore, to increase the number of physical qubits, we need denser integration of qubits and a larger qubit plane, which results in more frequent MBBE events and a larger anomaly size. 
According to the recent design of error-decoding units~\cite{holmes2020nisq+,ueno2021qecool,das2020scalable,das2021lilliput}, the available code distance is also limited by the processing power of classical peripherals. It is thus too optimistic to assume code distances can be increased to an arbitrarily large value. 
Finally, the increase of code distances means less throughput of logical operations, since the latency of most of the instructions is proportional to the code distances. 
Therefore, the MBBEs by cosmic rays are one of the main obstacles to achieving a scalable FTQC in the near future, and they should thus be mitigated to ensure the scalability of FTQC.

\subsection{Overview of Our Proposal: Q3DE}
\label{sec:overview}
\begin{figure*}
    \centering
    \includegraphics[width=1.0\textwidth]{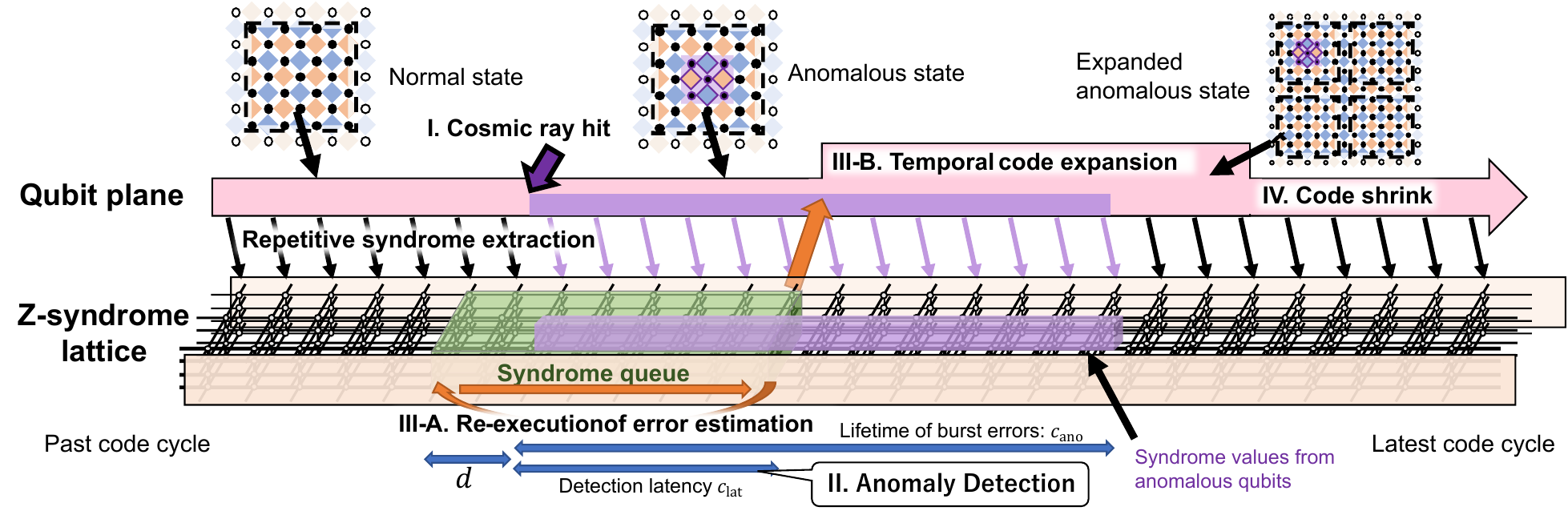}
    \caption{Overview of Q3DE. The timeline of interactions between the qubit plane and the syndrome lattice is shown.}
    \label{fig:concept_overview}
\end{figure*}

In this paper, we propose a solution to the problem of MBBEs by integrating several components and modifications into the standard design of FTQC.
There are three key components to our proposal: in-situ anomaly DEtection of MBBEs, dynamic code DEformation, and optimized error DEcoding, which we call {\it Q3DE}. Q3DE can be integrated by adding the components surrounded by the red dotted square in Fig.\,\ref{fig:architecture}. Its operational flow is shown in Fig.\,\ref{fig:concept_overview}.

The intuitive mechanism of Q3DE is as follows. I) A cosmic ray hits the substrate, and an MBBE occurs, which leads to high-error-rate physical qubits in an anomalous region. II) More active syndromes are generated from the region, and we can estimate the timing and position of MBBEs by checking their frequencies. We show in Sec.\,\ref{sec:anomaly_detection} that this detection can be implemented with a modest overhead. After the detection, we perform two reactions to mitigate the MBBE effect.
III-A) We temporally increase the code distance of the affected logical qubit, which can be achieved with a small latency by using code deformation techniques~\cite{bombin2009quantum}, as explained in Sec.\,\ref{sec:code_deformation}. III-B) The other action is decoder re-execution, which is explained in Sec.\,\ref{sec:error_decoding}. The degradation of logical error rates can be relaxed by re-estimating the most probable Pauli errors with the timing and position of MBBEs. IV) After the anomalous region vanishes, the code distance is reduced to the default value by using the code deformation techniques. 

\section{Detection of MBBE events}
\label{sec:anomaly_detection}
We first show that MBBEs, i.e., a temporal increase of physical error rates, can be detected by using the frequency of active syndrome nodes in the syndrome buffer. We provide statistical modeling of MBBEs for setting the confidence intervals of anomaly detection. Then, we present the procedure of anomaly detection.

We investigated two syndrome-counting strategies. One is simply counting the active syndromes in the buffer. The other is counting only for even-numbered cycles. While the first strategy seems natural, the confidence interval cannot be analytically derived due to statistical correlations. In contrast, while the confidence interval of the latter strategy can be calculated with the help of a central limit theorem~(CLT), this strategy is less efficient since it requires a double-sized window. Therefore, we derive the confidence interval of the latter strategy in Sec.\ref{subsec:anomaly_detection_theory}, and implement the procedure of the \texttt{anomaly detection unit} with the first strategy using the derived confidence intervals. We demonstrate that this strategy works with numerical simulation in Sec.\,\ref{sec:evaluation}. 

\subsection{Statistical modeling of syndrome sequence}
\label{subsec:anomaly_detection_theory}
We prove that the CLT is applicable to the time series of even-cycle syndrome values with small modification if physical errors on the qubit plane are independent, modeled as Pauli, acting on a single-qubit, and identical for every code cycle.\footnote{Even if these conditions are not met, we may be able to satisfy this assumption with a modest cost by introducing twirling techniques~\cite{gambetta2012characterization}.} 
Let $\hat{s}_{i,t}$ be a binomial random variable obtained as the outcome of the $i$-th stabilizer measurement at the $t$-th code cycle, and let $\hat{e}_{i,t}$ be a binomial random variable that becomes one if a physical Pauli error occurs at the $i$-th position and $t$-th cycle. Since the value of $\hat{s}_{i,t}$ is determined as the parity of the previous occurrence of Pauli errors, we can define a set of pairs of the time and position $E_{i,t}$ such that $\hat{s}_{i,t} =: \bigoplus_{(j,u) \in E_{i,t}} \hat{e}_{j,u}$.
Whether the syndrome node is active or not is determined by the parity of consecutive measurement outcomes, i.e., $\hat{v}_{i,t} = \hat{s}_{i,t} \oplus \hat{s}_{i,t-1} = \bigoplus_{(j,u) \in F_{i,t}} \hat{e}_{j,u}$, where $F_{i,t} := (E_{i,t} \cup E_{i,t-1}) \setminus (E_{i,t} \cap E_{i,t-1})$.
From the assumption of identical distributions over different cycles, the expectations and standard deviations of $\hat{v}_{i,t}$ are also identical\footnote{This does not always hold exactly, for example, at the first and last cycle of surface codes. Thus, this assumption holds except for such unusual cycles.} since $F_{i,t} \cap F_{i,u} = \emptyset$ for $|t-u| \ge 2$. We denote the expectations and standard deviations of $\hat{v}_{i,t}$ as $\mu$ and $\sigma$, respectively, which can be determined in the calibration process in advance.

We denote the number of active syndromes in the even-numbered cycles of the last $2 c_{\rm win}$ cycles as $\hat{V}_{i,t,c_{\rm win}} := \sum_{j=0}^{c_{\rm win}} \hat{v}_{i,t-2j}$, which is defined for $t > 2c_{\rm win}$. 
Since the random variable $\hat{V}_{i,t,c_{\rm win}}$ is identical, we can apply the CLT to $\hat{V}_{i,t,c_{\rm win}}$ and obtain
\begin{align}
\label{eq:normal_distribution_of_parity_count}
\hat{V}_{i,t,c_{\rm win}} \sim \mathcal{N}\left(c_{\rm win} \mu , c_{\rm win} \sigma^2\right).
\end{align}
Thus, when the window size $c_{\rm win}$ is sufficiently large, the distribution of the count of active syndromes in the window can be approximated by a normal distribution.
Suppose that the normal distribution approximation introduced in Eq.\,(\ref{eq:normal_distribution_of_parity_count}) is valid.
When an MBBE does not occur, we can expect that, with a confidence level $1-\alpha$, the count of active syndromes $\hat{V}_{i,t,c_{\rm win}}$ satisfies $\hat{V}_{i,t,c_{\rm win}} < \hat{V}_{\rm th}$, where
\begin{align}
\label{eq:anomaly_detection_criterion}
\hat{V}_{\rm th} := c_{\rm win} \mu + \sqrt{2 c_{\rm win} \sigma^2} \, \mathrm{erf}^{-1}(1-\alpha),
\end{align}
and $\mathrm{erf}^{-1}$ is the inverse of the Gauss error function.
Conversely, an observation of $\hat{V}_{i,t,c_{\rm win}}$ not satisfying Eq.\,(\ref{eq:anomaly_detection_criterion}) implies that physical qubits around the position $i$ have dropped into an anomalous state. 
Supposing an anomaly size $d_{\rm ano}$, about $d_{\rm ano}^2$ active syndrome counters would be detected as anomalous states. We detect the occurrence of MBBEs by comparing the number of counters above the confidence interval and an integer $n_{\rm th}$. Since it is enough for FTQC to suppress the rates of false-positive and true-negative anomaly detection to a value smaller than a logical error rate $p_{\rm L}$, we can roughly set a criterion as $\frac{\ln p_{\rm L}}{\ln \alpha} < n_{\rm th} < d_{\rm ano}^2 - \frac{\ln p_{\rm L}}{\ln \alpha}$, while we expect the actual $n_{\rm th}$ should be determined according to experiments. Note that when there is no $n_{\rm th}$ that satisfies the inequality with reasonable $\alpha$, this implies that physical qubits are already tolerant to MBBEs, and logical error rates are not significantly degraded.

\subsection{Procedure of anomaly detection unit}
\label{subsec:anomaly_detection_procedure}
In accordance with the theoretical modeling in the last section, we describe the procedure of the \texttt{anomaly detection unit}. Let sampled $m$-bit nodes obtained in the $t$-th code cycle be $\vec{v}_{t} := (v_{1, t}, \dots, v_{m, t})$, where the $i$-th element of $\vec{v}_{t}$ becomes one when the $i$-th syndrome node becomes active at the $t$-th code cycle. 
The \texttt{anomaly detection unit} keeps the number of active nodes in the latest $c_{\rm win}$-length window for each position in the \texttt{active node counter}. In other words, the \texttt{active node counter} contains a list of integers $\vec{V}_t$ where the $i$-th element is $\vec{V}_t[i] := \sum_{u=0}^{c_{\rm win}} \vec{v}_{t-u} [i]$. To update $\vec{V}_{t+1}$ from $\vec{V}_t$, the \texttt{anomaly detection unit} fetches $\vec{v}_{t+1}$ and $\vec{v}_{t - c_{\rm win}}$ and updates as $\vec{V}_{t+1} \leftarrow \vec{V}_{t} + \vec{v}_{t+1} - \vec{v}_{t - c_{\rm win}}$ for each element. Thus, we need a syndrome queue larger than $c_{\rm win}$. 
After updating the active syndrome counter, the \texttt{anomaly detection unit} checks the number of positions where the count is larger than a threshold $V_{\rm th}$. Suppose that the number of counts above the threshold is $n_{\rm ano}$. Then, the \texttt{anomaly detection unit} determines that there is an MBBE if $n_{\rm ano}$ becomes larger than a threshold count $n_{\rm th}$.
Once we detect MBBEs, their timing can be estimated from the size of the detection window $c_{\rm win}$, which is numerically evaluated in Sec.\,\ref{sec:evaluation}. We also estimate the position of the anomalous region as the median of detected anomalous-qubit positions. Then, we temporally remove the detected positions around the median from the count of $n_{\rm ano}$ for the lifetime of MBBEs and continue the anomaly detection. If two MBBE events occur at almost the same cycle, the second one is detected immediately after the first detection.

\section{Temporal code expansion}
The code distance of logical qubits affected by MBBEs is dynamically increased immediately after the detection. We introduce low-overhead implementation of the dynamical expansion of code distances. 

\label{sec:code_deformation}
\subsection{Code expansion by code deformation}
\label{subsec:code_expansion_scheme}
\begin{figure}
    \centering
    \includegraphics[width=0.5\textwidth]{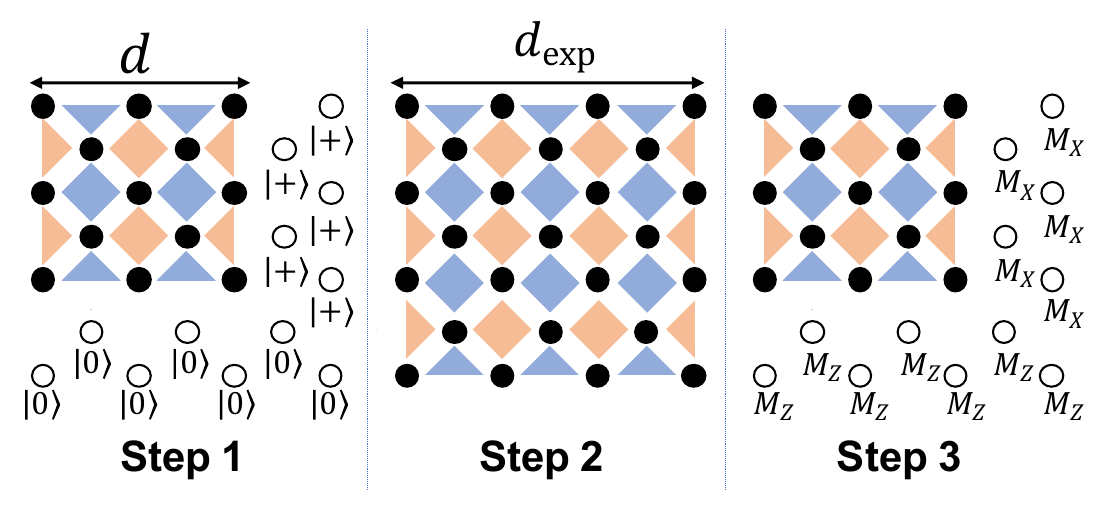}
    \caption{Three steps for \texttt{op\_expand} instruction. These figures show a procedure to expand the code distance of a logical qubit from $d$ to $d_{\rm exp}$ and shrink it from $d_{\rm exp}$ to $d$.}
    \label{fig:code_deformation}
\end{figure}

Code distances can be changed during the computation with a technique known as code deformation~\cite{bombin2009quantum}. Figure~\ref{fig:code_deformation} shows its procedure.
The code deformation is performed with unused data qubits indicated by white circles. First, we initialize unused data qubits to $\ket{0}$ or $\ket{+}$~(step~1). Then, we update the \texttt{stabilizer map} to perform a new set of stabilizer measurements as an expanded surface code~(step~2) and keep the expanded code distance until MBBEs finish. To shrink a code distance, we perform Pauli-$X,Z$ measurements on qubits used for the expansion~(step~3) and update the \texttt{stabilizer map} again to perform stabilizer measurements of the original pattern.

Note that there are several other methods to fault-tolerantly mitigate MBBEs according to the information of them, and the best choice relies on the policy of qubit allocations. We can choose several other strategies, e.g., moving affected logical qubits to another space or expanding code distance by default and prohibiting the shrink under the existence of anomalous regions. The comparison between these methods is left as future work.

\subsection{Procedure of temporal code expansion}
When the \texttt{anomaly detection unit} judges that a logical qubit is affected by MBBEs, it inserts an instruction \texttt{op\_expand} to the \texttt{expansion queue}.
This instruction makes the code distance of an affected logical qubit temporally increase from $d$ to $d_{\rm exp}$ as soon as possible and keeps it for a typical lifetime of MBBEs. 
As analyzed in the next section, since the effect of MBBEs can be treated as the effective reduction of code distance by $2d_{\rm ano}$, it is enough to set the expanded code distance $d_{\rm exp}$ as a value larger than $d+2d_{\rm ano}$.
Since $2d_{\rm ano} \ll d$ suffices in practice, doubling the code distance using $2 \times 2$ surface-code blocks for a logical qubit is enough for decreasing the logical error rates to an original value. When \texttt{op\_expand} is executed on a region where it is already expanded, we increase the keeping time at step~2.

\section{Re-execution of error decoding}
\label{sec:error_decoding}
The other reaction to the anomaly detection is decoder re-execution. 
We provide a concrete 2D-syndrome-layer example for showing intuitive reasons why the re-execution is effective in Fig.\,\ref{fig:effective_code_distance}~(a). When qubits in the purple area are anomalous but the decoder does not know it, the upper matching with five edges is estimated. If the decoder knows that qubits in the purple area suffer from large physical error rates, the lower solution is more probable, although it uses more edges than the upper one. This means the lower solution can retrieve the original logical states with a higher probability.
This example clearly shows that the knowledge of MBBEs enables higher-performance decoding algorithms, which motivates us to switch the decoding algorithms to suppress logical error rates adaptively. Moreover, this technique can be applied to the estimation in the past cycles by rolling back the decoding process. This is a key advantage since the code-expansion technique can protect logical qubits after the latency of the MBBE detection.

In this section, we first provide an approximated analysis to determine the advantage of decoder re-execution. Next, we show that an MBBE-aware matching algorithm can be implemented with modest changes to standard error-decoding strategies. Finally, we show how to roll back and switch the decoding process with a modest overhead.

\subsection{First-order analysis of decoder re-execution}
\label{subsec:decoder_analysis}
\begin{figure*}
    \centering
    \includegraphics[width=1.0\textwidth]{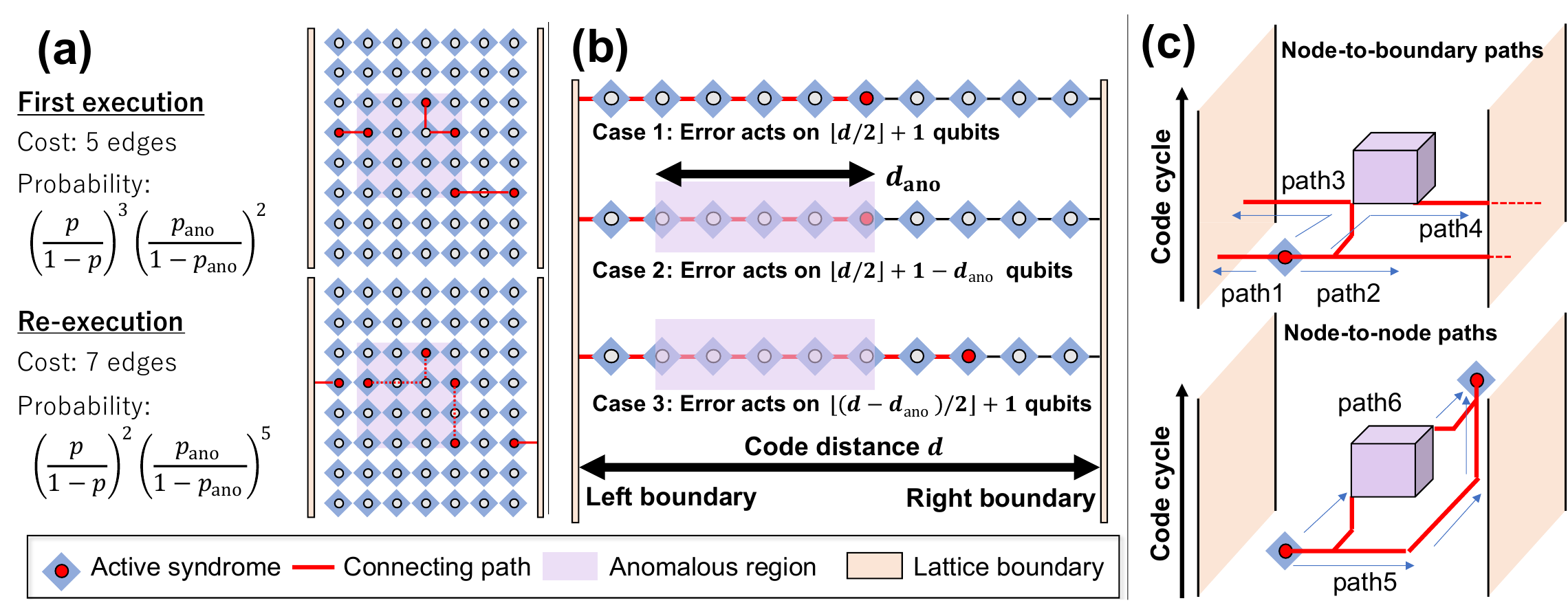}
    \caption{Schematic diagrams showing the mechanism and analysis of decoder re-execution. We only show a $Z$-lattice for readability, but the same property holds for the $X$-lattice. (a) A working example where decoder re-execution performs better estimation. (b) Schematic diagram showing the minimum number of required flips of normal edges to incur logical errors. (c) A list of possible node-to-boundary and node-to-node paths in the matching procedure. One among these six paths is the shortest to match nodes, and thus the computational overheads of greedy-based optimized decoding are constant.}
    \label{fig:effective_code_distance}
\end{figure*}

We evaluate the performance of surface codes with an anomalous region and the improvement by decoder re-execution with a first-order approximation. In this approximation, we assume that only the paths with the lowest order of physical error rates of normal qubits $p$ contribute to the logical error rates. In other words, we count the minimum number of normal edges that must be flipped to incur a wrong decision of the decoding unit, as shown in Fig.\,\ref{fig:effective_code_distance}~(b).
This approximation is justified when $p$ is small and $d$ is large, which provides us with the performance of QEC codes that is difficult to evaluate with numerical analysis. It also provides insight for understanding the effect of MBBEs for large code distances and small physical error rates. 

When there is no anomalous region, Pauli errors acting on more than half of the code distance $\lfloor d/2 \rfloor+1$ are required to incur a logical error (Case~1). Thus, the logical error rate $p_{\rm L}$ is known to scale as a function of code distance $p_{\rm L}(d) \propto (p/p_{\rm th})^{\lfloor d/2 \rfloor+1}$~\cite{fowler2012towards}, where $p_{\rm th}$ is a constant.
Meanwhile, when there exists an anomalous region with size $d_{\rm ano}$, Pauli errors acting only on $(\lfloor d/2 \rfloor+1-d_{\rm ano})$ normal qubits are required~(Case~2).
This implies that the logical error rate with an anomalous region scales as $p_{\rm L, ano}(d) \propto (p/p_{\rm th})^{\lfloor d/2 \rfloor+1-d_{\rm ano}}$, which is equal to the reduction of the code distance from $d$ to $d-2d_{\rm ano}$. 
If the decoding unit knows the position of the anomalous region, we choose the most probable matching that eliminates the edges in the region. Thus, $(\lfloor (d - d_{\rm ano})/2 \rfloor + 1)$-qubit Pauli error on normal qubits is required to induce a wrong decision of the decoding unit~(Case~3), which corresponds to the reduction of code distances from $d$ to $d-d_{\rm ano}$.
This analysis implies that the threshold of surface codes does not change by MBBEs and decoder re-execution, which agrees with the results in Fig.\,\ref{fig:logical_error_prob_motivation}.

\subsection{Algorithm for weighted matching}
\label{subsec:decoder_nonuniform}
In order to take the MBBEs into account in the decoding process, we need to solve matching problems for non-uniform weights, where the weight of edges in the anomalous regions is set as $-\log \frac{p_{\rm ano}}{1-p_{\rm ano}}$ and that of normal qubits as $-\log \frac{p}{1-p}$. 
We can find the most probable Pauli errors on non-uniform weighted graphs by enumerating the shortest path between active nodes with the Dijkstra algorithm and by finding an MWPM with Edmonds' blossom algorithm~\cite{edmonds1965paths}.
While this approach provides the exact solution, the procedure takes a longer time than a $1 {\rm \mu s}$ code cycle even in the case of uniform weights. Fast approximated algorithms have therefore been investigated~\cite{das2020scalable,holmes2020nisq+,ueno2021qecool,das2021lilliput}. One of the fastest approximated algorithms for uniform-weight graphs is the greedy-based algorithm~\cite{holmes2020nisq+,ueno2021qecool}. In this algorithm, we set $i=1$ and match active nodes that can be connected with a path shorter than $i$ in a greedy manner. We iterate this procedure from $i=1$ to $d$. The length between two nodes can be calculated as the Manhattan distance. 
When the non-uniformity is caused by a single anomalous region, we can extend this algorithm for weighted graphs with a small modification: namely, we replace the calculation method of the shortest path between two given nodes with a method that chooses the shortest path among candidates, as shown in Fig.\,\ref{fig:effective_code_distance}~(c). Since this diagnosis can be done in parallel and takes a constant time, this modification incurs only a small overhead on the throughput, as verified with logic synthesis in Sec.\,\ref{subsec:scalability_pipeline_latency}.
Note that the physical error rate inside an anomalous region is actually not uniform, and it decays according to the elapsed time and the position from the center. Nevertheless, we can introduce this concept approximately by letting the inside of the contour for $p_{\rm ano}$ be an anomalous region.

We also note the feasibility of implementing the existing error-estimation algorithm on the weighted graph. For the union-find-based strategy~\cite{delfosse2017almost,das2020scalable}, Pattison~\textit{et al.}~\cite{pattison2021improved} proposed a low-overhead algorithm to execute the union-find decoder on weighted graphs. The existing greedy-based strategy~\cite{holmes2020nisq+,ueno2021qecool} is implemented with single-flux-quantum~(SFQ) circuits~\cite{holmes2020nisq+,ueno2021qecool}, so introducing non-uniformity to these implementations is non-trivial. We anticipate that this concept can be introduced by switching spike propagation modes corresponding to normal and anomalous regions, but its implementation is left to future work.

\subsection{Procedure for decoder re-execution}
\label{subsec:decoder_reexecution}
In this subsection, we discuss how to roll back the past estimation of recovery Pauli operations and switch the decoding strategy.
Suppose that the \texttt{anomaly detection unit} detects an anomalous region at the $t$-th cycle with a latency of $c_{\rm lat}$ cycles. Then, the timing of the occurrence of the anomalous region is the $(t-c_{\rm lat})$-th cycle. Thus, the states of the \texttt{syndrome queue}, \texttt{decoding unit}, \texttt{Pauli frame}, and \texttt{classical register} should be rolled back to those at the $(t-c_{\rm lat}-d)$-th cycle.
If a \texttt{read} instruction has already referred to any entry corrected after the $(t-c_{\rm lat}-d)$-th cycle, the rollback procedure is aborted since it also needs to roll back the state of the host CPU, which is expected to be too costly.
The most naive implementation of the rollback is to keep all the snapshots of relevant units for recent $c_{\rm lat}+d$ cycles and load the oldest snapshot. After the rollback, the decoding unit restarts the procedure with weighted edges. While this method is simple, it consumes a large amount of memory. Thus, we show a more efficient method for the rollback.
The rollback of the \texttt{syndrome queue} and \texttt{decoding unit} can be implemented as follows. We enlarge the window size of \texttt{syndrome queue} to $c_{\rm lat}+d$ cycles and keep them even if they are matched. When we roll back the units, we forget the matches of active nodes in recent $c_{\rm lat}+d$ cycles, reset the decoding units, and restart the decoding.
Since all the operations on the \texttt{Pauli frame} and \texttt{classical register} are reversible, we can revert them by storing the update operations for them.
The output of the \texttt{decoding unit} required for the rollback of the \texttt{Pauli frame} is stored in the \texttt{matching queue}. Since the full records of matching results consume a large memory space, we reduce the size of \texttt{matching queue} by taking the sum of each $c_{\rm bat}$ cycle with the information of pairs connecting neighboring batches. This technique reduces the size of the \texttt{matching queue} by $c_{\rm bat}$ times but requires additional $c_{\rm bat}$-layer memory space for the \texttt{syndrome queue}. We can find that setting $c_{\rm bat} = \sqrt{2c_{\rm win}}$ will minimize the total amount of buffer memory. See Sec.\,\ref{subsec:scalability_memory_space} for the concrete evaluation.
Also, since the \texttt{Pauli frame} must be updated according to the execution of logical instructions~(see Sec.\,\ref{sec:background}), its update history is also stored in the \texttt{instruction history buffer}. The rollback of the \texttt{classical register} can be achieved simply by marking the entries measured after the $(t-c_{\rm lat}-d)$-th cycle as ``not-error-corrected''.

\section{Numerical evaluation}
\label{sec:evaluation}

\subsection{Numerical simulation settings}
\label{subsec:evaluation_simulation_settings}
In this section, we numerically evaluate the detection errors and latency of anomaly detection and the logical error reduction by the decoder re-execution. To efficiently capture the performance of QEC strategies, we assume the following for the noise models; Stochastic Pauli errors occur on data and ancillary qubit at the beginning of each code cycle. In each error position, Pauli-$X,Y,Z$ errors occur with probability $p/2$ for normal qubits and $p_{\rm ano}/2$ for anomalous qubits. 
1) 
The settings are as follows.
1) Noise maps are inserted at the beginning of each code cycle on data and ancillary qubits. 
2) In each noise map, Pauli-$X, Y, Z$ errors occur with probability $p/2$ for normal qubits and $p_{\rm ano}/2$ for anomalous qubits. 
3) Logical error rates are evaluated as a logical Pauli-$X$ error rate per cycle in the procedure of $d$-cycle idling.
4) Decoding units ignore correlations due to Pauli-$Y$ errors and estimate the occurrence of Pauli-$X$ and $Z$ errors independently.
5) We evaluate logical error rates using the Monte-Carlo method with at least $10^5$ samples for each data point.

These are popular methods and assumptions to efficiently capture the performance of error-correction strategies~\cite{ueno2021qecool,holmes2020nisq+,duckering2020virtualized}. We implemented our own error simulators for all simulations except the error estimation, for which we used Kolmogorov's implementation of Edmonds' blossom algorithm~\cite{edmonds1965paths,kolmogorov2009blossom}. We verified that the results without considering MBBEs agree with the existing results such as Ref.\cite{ueno2021qecool}.

\subsection{Anomaly detection unit}
\begin{figure}
    \centering
    \includegraphics[width=0.5\textwidth]{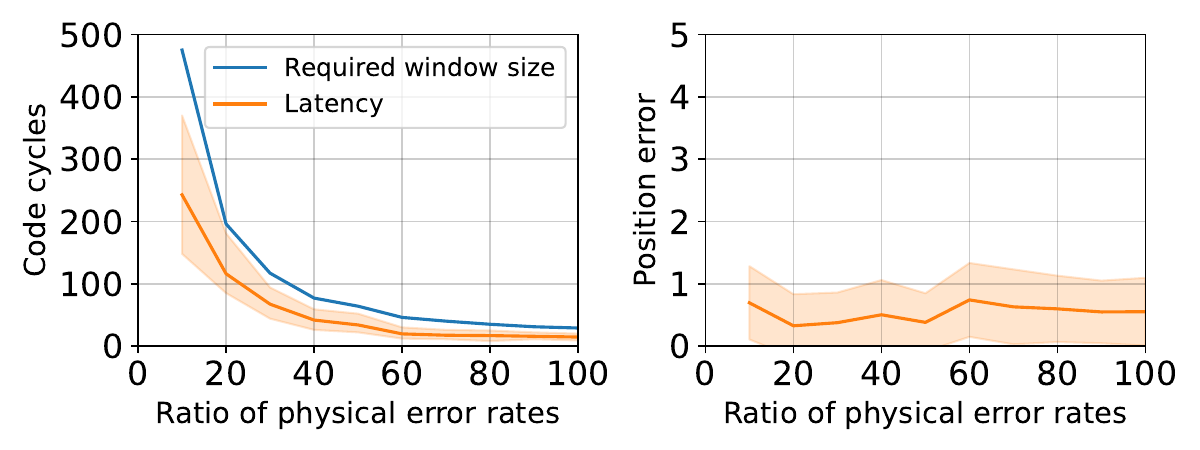}
    \caption{Numerical evaluation of the \texttt{anomaly detection unit}. 
    (Left) Required window size $c_{\rm win}$ for 1\% detection error rates and detection latency. (Right) The error of estimated positions of anomalous regions.
    }
    \label{fig:anomaly_detection_evaluation}
\end{figure}

We numerically simulated the behavior of the \texttt{anomaly detection units} to clarify the desirable settings according to experimental parameters. There are three parameters that determine the performance of the \texttt{anomaly detection units}: the size of window $c_{\rm win}$, the confidence level $1-\alpha$, and the threshold number of anomalous qubits $n_{\rm th}$. 
In the following evaluation, we set $1-\alpha = 0.99$. We assume that $\mu$ and $\sigma$, i.e., the frequency of active nodes and their variance of normal qubits, are known in the pre-calibration phase. Therefore, the main interest of the numerical evaluation is to determine $c_{\rm win}$ and $n_{\rm th}$ for realistic parameters. 
To determine them, we simulate the anomaly detection process. The results are shown in Fig.\,\ref{fig:anomaly_detection_evaluation}, where the blue line in the left graph shows the minimum $c_{\rm win}$ required for achieving the probabilities of the false-positive and true-negative detection of anomalous qubits being below 1\% under $p=10^{-3}$ according to the temporal increase ratio of physical error rates $p_{\rm ano}/p$.
With the selected $c_{\rm win}$, we heuristically choose $n_{\rm th}=20$ and evaluate the latency of the anomaly detection and the error of the estimated positions of anomalous regions for $d_{\rm ano} = 4$ and $d=21$, which are plotted as the orange line in the left and right graphs, respectively. 
We can see that the timing and position are estimated accurately. Since the code cycle is about $1~\mathrm{\mu s}$ and the MBBEs last a few tens of milliseconds, the period for which logical qubits are exposed to MBBEs without any reaction becomes about $10^{-3}$ times shorter.

\subsection{Decoder re-execution}
\label{subsec:evaluation_optimized_decoding}
\begin{figure}
    \centering
    \includegraphics[width=0.5\textwidth]{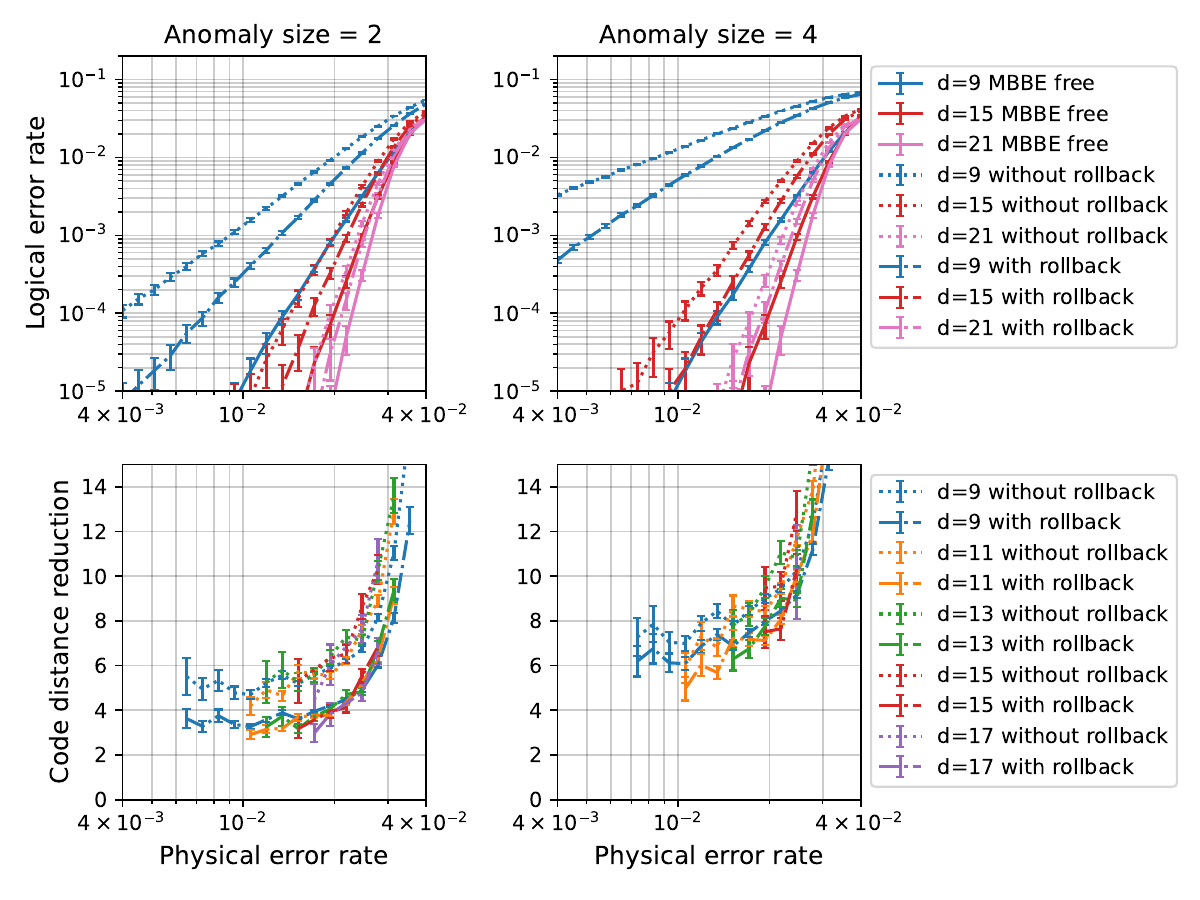}
    \caption{Logical error rates and effective reductions of code distances plotted with anomaly size $d_{\rm ano}=2$ (left two graphs) and $d_{\rm ano}=4$ (right two graphs).}
    \label{fig:logical_error_prob_rollback}
\end{figure}

We evaluated the improvement of logical error rates by the decoder re-execution. The top two graphs of Fig.\,\ref{fig:logical_error_prob_rollback} show the logical error rate per code cycle with and without rollback techniques for $d_{\rm ano}=2$ and $4$. The solid lines represent logical error rates without an MBBE. The broken and dotted lines represent those under an MBBE with and without the rollback, respectively. 
We observed that the rollback techniques improve the logical error rates, and the improvement ratio becomes significant as physical error rates become small.

Next, we evaluated the effective reduction of code distances by MBBEs with the following equation.
\begin{align}
d-d_{\rm eff} = \left(\ln \frac{p_{\rm L, ano}(p, d)}{p_{\rm L}(p,d)}\right) \bigg/ \left(\frac{1}{2} \ln \frac{p_{\rm L}(p, d-2)}{p_{\rm L}(p, d)}\right)
\end{align}
The reduction of code distances is plotted in the bottom two graphs of Fig.\,\ref{fig:logical_error_prob_rollback}. Since the effective code distances have large uncertainty, we only plot the data points of which the standard error is smaller than four, and we used the standard errors divided by four as error bars for visibility. 
According to the first-order analysis in Sec.\,\ref{sec:error_decoding}, the reduction converges to $2d_{\rm ano}$ without rollback and to $d_{\rm ano}$ with rollback for sufficiently small $p$ and large $d$. 
While it is difficult to observe clear convergence in the case of $d_{\rm ano}=4$ due to large standard errors, we can observe the trend of convergence for $d_{\rm ano}=2$. 
The converged values for $d_{\rm ano}=2$ seem slightly larger than the expected values. We expect this discrepancy occurs because the first-order analysis is not sufficiently justified in a region of simulated physical error rates and code distances or due to the coefficient stemming from the number of minimum-weight paths in the first-order approximation. Note that since smaller physical error rates and larger code distances than those of simulations are required to execute long-term applications~\cite{babbush2018encoding,kivlichan2020improved,gidney2021factor}, e.g., $p=0.001$ and $d = 21$, we expect the first-order analysis is justified in practical cases.

\section{Scalability discussion}
\label{sec:discussion}
In this section, we discuss the scalability of FTQC with and without Q3DE and show that Q3DE reduces the difficulties in the future scaling of FTQC. For the comparison, we utilize an FTQC architecture that mitigates MBBEs by increasing the default code distance as the baseline.

\subsection{Qubit count}
\label{subsec:scalability_qubit_count}
\begin{figure}
    \centering
    \includegraphics[width=0.5\textwidth]{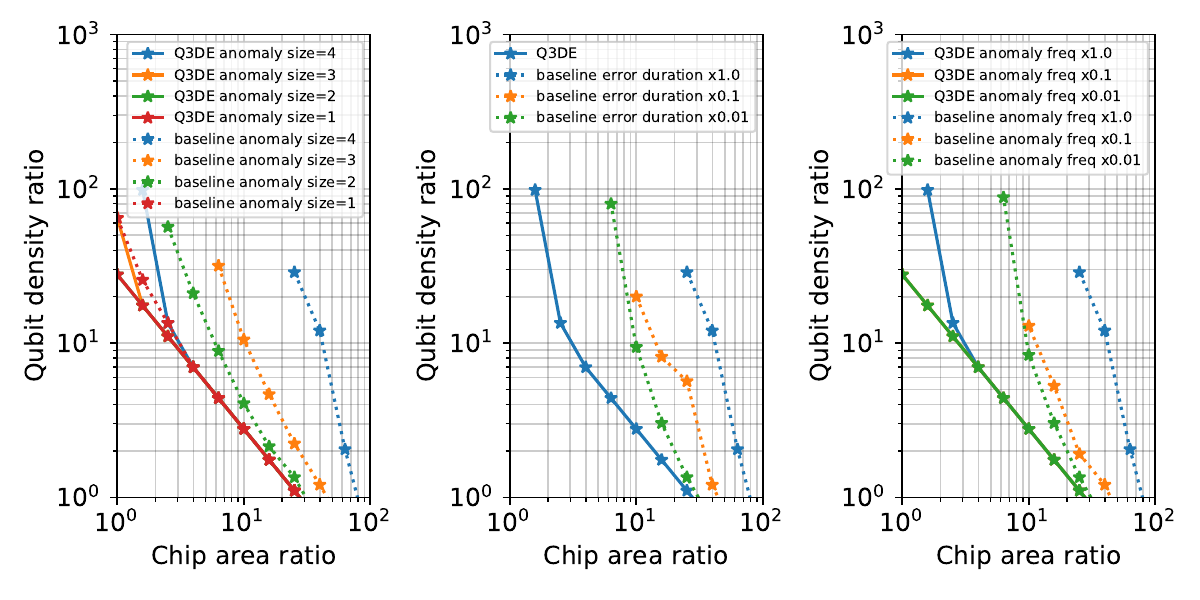}
    \caption{The required area and qubit density of a qubit plane per logical qubit to achieve a logical error rate below $10^{-10}$ are plotted as a double logarithmic graph.}
    \label{fig:scalability_estimate}
\end{figure}

To achieve a logical error rate with given physical error rates, we need to increase the number of qubits on the qubit plane, which means increasing the chip area and qubit density. Figure\,\ref{fig:scalability_estimate} shows the required chip area and qubit density per logical qubit relative to the values of Google's Sycamore chip~\cite{mcewen2021resolving} to achieve a logical error rate below $10^{-10}$.
We calculated the required qubit density for a given chip area as follows. We set the physical error probability over the threshold value $p/p_{\rm th}$ as $0.1$ and code cycle as $1~{\rm \mu s}$. We choose $d_{\rm ano}=4$, $f_{\rm ano}=0.1$, and $\tau_{\rm ano}=25 {\rm ms}$ as the baseline, and change each parameter to smaller values to check the performance of Q3DE in several parameter regimes.
When we swept the chip area and qubit density, we assumed that the frequency of cosmic rays and the size of anomalous regions increase linearly to them, respectively. 
For each parameter setting, we put anomalous regions at random positions with the Poisson distribution and remove them after an error duration. When the smallest number of normal edges in logical Pauli errors is reduced from $d$ to $d-c$, we assumed that the effective code distance becomes $d-c$ with Q3DE and $d-2c$ without Q3DE according to the discussion of Sec.\,\ref{sec:error_decoding}, and calculate logical error rates with $p_{\rm L}(d) = 0.1(p/p_{\rm th})^{\lfloor (d_{\rm eff}+1)/2 \rfloor}$. We simulate this process for $10^8$ code cycles and calculate the averaged logical error rate. We also assumed that the Q3DE reduces the period exposed to MBBEs to the latency of anomaly detection $c_{\rm lat}=30$ since it expands code distances to a sufficiently large value.
Starting this evaluation from $d=11$, we increase the code distance until the logical error rates become below $10^{-1}$ by increasing the qubit density and find the minimum required qubit density. 

When qubit planes do not suffer from cosmic rays, the required qubit density is proportional to the inverse of the chip area. On the other hand, when we consider MBBE effects, the required qubit density becomes large. Comparing the performance with and without the Q3DE architecture, we can see that the Q3DE can suppress the MBBE effect without significantly increasing physical qubits. In particular, when the qubit density ratio is about ten, the reduction of qubit count is up to about ten times in the baseline settings.
Thus, the Q3DE is indispensable not only for efficiently using qubit planes but also when the achievable area and density per logical qubit are limited.

\subsection{Instruction throughput}
The adaptive insertion of \texttt{op\_expand} by dynamical code expansion may block the following operations since it consumes unused qubit space on the qubit plane and blocks the following instructions such as \texttt{ZZ\_meas} (i.e., lattice surgery, which connects logical qubits using unused blocks). The baseline method, i.e., the increase of the default code distance, also reduces the instruction throughput, since the latency of several instructions is proportional to the code distances.

\begin{figure}
    \begin{minipage}{0.4\linewidth}
        \centering
        \includegraphics[width=0.8\linewidth]{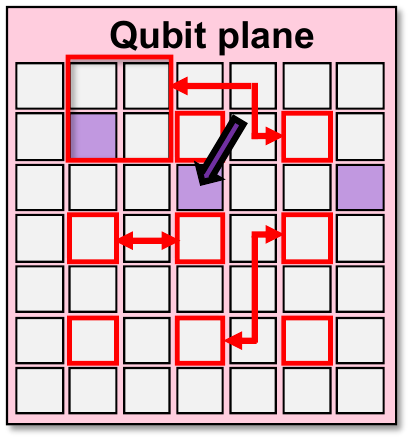}
    \end{minipage}
    \begin{minipage}{0.5\linewidth}
        \centering
        \includegraphics[width=1\linewidth]{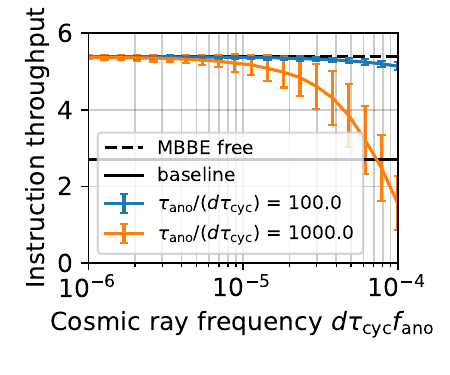}
    \end{minipage}
    \caption{(Left) An example of two-qubit logical operations on a $7 \times 7$ qubit plane.  (Right) Instruction throughput with and without Q3DE under cosmic rays.}
    \label{fig:scalability_throughput}
\end{figure}

We compared the instruction throughput by simulating $10^4$ \texttt{meas\_ZZ} instructions acting on two randomly chosen logical qubits. We performed numerical simulation with a qubit plane consisting of 25 logical qubits allocated on an $11 \times 11$ grid.
The average number of completed instructions during $d$ code cycles is shown in Fig.\,\ref{fig:scalability_throughput} with a schematic diagram of the logical operations. In this simulation, we assumed that the MBBEs on unused blocks are detected via direct measurements of data qubits and the instruction scheduler avoids using these blocks. We assumed that an MBBE occur on each block with probability $d \tau_{\rm cyc} f_{\rm ano}$ in each $d$ cycle and lasts for $100d$ or $1000d$ cycles. We also assumed that the code distance is doubled in the code expansion of Q3DE and in the baseline solution. We scheduled logical operations with a greedy algorithm.
Since an MBBE frequency per $d$ code cycle is about $d \tau_{\rm cyc} f_{\rm ano} \sim 10^{-5}$ in realistic parameters~\cite{mcewen2021resolving}, the degradation of throughput is acceptable and better than the baseline. If the frequencies and period of MBBEs are improved in the future, the throughput is doubled compared to the baseline.

Decoder re-execution may also degrade instruction throughput, since it rolls back the processing of the decoding unit in accordance with MBBEs, the rollback delays the correction of corresponding classical registers, and the execution of the \texttt{read} operation may be blocked. The most pessimistic scenario is that a \texttt{read} operation is executed just after every rollback, and all the consecutive instructions are blocked by this \texttt{read} operation. In that case, all the following instructions need to wait for the decoding unit to recalculate the matching of $d+c_{\rm lat}$ cycles. Since the instruction needs to wait for matching $d$ cycles without rollback, the execution cycles of the \texttt{read} instruction get up to $1 + \frac{c_{\rm lat}}{d}$ times larger, which is about two or three in realistic parameters. Since such a pessimistic situation would be rare, we believe this overhead is negligible in the total computation time.

\subsection{Overhead of syndrome buffer sizes}
\label{subsec:scalability_memory_space}
We list the additional memory space for the decoder re-execution in Table\,\ref{tab:memory_overhead}.
\begin{table}[t]
  \centering
  \caption{Summary of memory overheads of Q3DE.}
  \label{tab:memory_overhead}
  \scalebox{1.0}{\begin{tabular}{|l|c|c|} 
    \hline
    Unit & Order & Size \\ 
    \hline \hline
    syndrome queue & $2 d^2 (c_{\rm win}+\sqrt{2c_{\rm win}})$ & 623~kbit\\
    active node counter & $2 d^2 \log_2 c_{\rm win}$ & 16~kbit \\
    matching queue & $2 d^2 \sqrt{c_{\rm win}/2}$ & 24~kbit \\ 
    inst. hist. buffer & negligible & -- \\ 
    expansion queue & negligible & -- \\ 
    \hline
  \end{tabular}}
\end{table}
The right-most column shows the memory sizes per logical qubit with $d=31$, $p=10^{-3}$, and $c_{\rm win}=300$. These are sufficient for large-scale applications and $c_{\rm win}$ is pessimistically chosen from Fig.\,\ref{fig:anomaly_detection_evaluation}. The dominant overhead is the syndrome queue, which is about ten times larger than the MBBE-free case $2d^3 \sim 58~{\rm kbit}$. If $c_{\rm win}$ is comparable to the code distance $d$, the overhead becomes almost negligible. Thus, we presume these sizes are acceptable in current technologies. 

\subsection{Overhead of decoding unit}
\label{subsec:scalability_pipeline_latency}
Q3DE imposes an additional task in the post-processing pipeline that has to satisfy the time constraint identical to the code cycle (i.e., about $1~{\rm \mu s}$) on average.
Since additional queues and counters can be updated in parallel with the decoding unit, and since their updates can be performed with simple arithmetic operations, these additional processes do not conflict with the allowable execution time constraint when MBBEs do not occur. 
The degradation of matching speed in the \texttt{decoding units} by introducing the MBBE-aware matching algorithms may create a bottleneck. 
To quantify the overhead, we implemented the greedy-based decoder in QECOOL~\cite{ueno2021qecool} with~(\textit{Q3DE}) and without the modification~(\textit{BASE}) shown in Sec.\,\ref{subsec:decoder_nonuniform}.
Instead of the SFQ circuit design in QECOOL, we designed FPGA circuits using high-level synthesis to evaluate the overheads of Q3DE. The high-level~(C++) description of circuits are verified in function-level simulations using randomly generated test patterns.

In each code cycle, the positions and distances to the anomaly and boundary are extracted and pushed to the active nodes queue~(\textit{ANQ}) in the decoding unit.
The unit calculates all-to-all shortest paths between ANQ entries in parallel and with pipelining, picks the shortest pair, and sends it to the Pauli frame and matching queue. 
The critical factor in this implementation is the entry size of ANQ: 1) it must be large enough that a buffer overflow occurs with less probability than the logical error rate, and 2) to satisfy the execution time constraint, the number of path evaluation and comparison circuits has to increase along with the growth in entry size. 
To satisfy the first criterion, $30$ entries are estimated to be enough for $p=10^{-4}, d=15, p_{\rm L}=10^{-15}$ and $70$ entries for $p=10^{-3}, d=31, p_{\rm L}=10^{-15}$.
\begin{table}[t]
  \centering
  \caption{FPGA implementation of greedy-based decoder.}
  \label{tab:synthesis}
  \scalebox{1.0}{\begin{tabular}{|l|c|c|c|} 
    \hline
    Configuration & FF (\%) & LUT (\%) & throughput\\ 
    \hline \hline
    40 -- BASE &  8,991 ( 4) & 14,679 ( 6) & 4.66 \\
    40 -- Q3DE & 13,855 ( 6) & 20,279 ( 9) & 4.25 \\
    80 -- BASE & 13,211 ( 6) & 36,668 (16) & 1.81 \\
    80 -- Q3DE & 22,751 (10) & 54,638 (24) & 1.79 \\
    \hline
  \end{tabular}}
\end{table}
Thus, we compared the post-layout designs for several parameters using \texttt{Vitis HLS 2021.2}.
The operating frequency is $400~{\rm MHz}$ and the target device is \texttt{Zynq} \texttt{Ultrascale+} \texttt{XCZU7EV-2FFVC1156} \texttt{MPSoC}~(\texttt{ZCU104} evaluation board).
The results are listed in Table~\,\ref{tab:synthesis} where configuration denotes ``ANQ Entry size -- BASE/Q3DE'' and throughput is shown in ``match/$\mu$s''. 
The throughput should be faster than the average number of active nodes per code cycle.
We can see that the slow-down of the MBBE-aware matching is acceptable.
While the frequencies of active nodes may temporally increase during MBBEs, it is acceptable if the matching speed of decoding units averaged with MBBE-free cycles is sufficiently fast.

The gap in resource utilization mainly comes from the difference in the bit-width in the calculations. Designs employ 8-bit and 16-bit unsigned integers for path length evaluations in BASE and Q3DE, respectively. Another reason is the increase of paths taken into account to consider anomalous regions. However, the area overhead of Q3DE is low enough to fit an embedded system class FPGA.

\section{Applicable scope of Q3DE}
\label{sec:applicability}
In previous sections, we have discussed the application of Q3DE to cosmic-ray-induced MBBEs based on the state-of-art superconducting quantum chip.
However, the property of MBBEs varies with the qubit technique advancing, e.g., device-level improvement or development of novel quantum devices.
We will verify the following basic assumptions in the Q3DE in the technology trend: 
1) There are non-negligible MBBE events that incur high-error-rate qubits. 2) Logical qubits are encoded with topological stabilizer codes that support the temporal expansion of code distances. 3) There are vacant blocks around logical qubits for operations.
The assumptions 2) and 3) are also assumed in a wide range of FTQC proposals~\cite{fu2018microarchitecture,holmes2020nisq+,fu2019control,fu2019eqasm,duckering2020virtualized}. 
To show that assumption 1) is reasonable in the future, in this section, we discuss the futuristic parameters of MBBEs and other factors inducing MBBEs, and show that the idea of Q3DE can be used as a versatile method to reduce the effect of MBBE events.

\subsection{Cosmic-ray induced MBBEs}
While this paper focuses on the parameters observed by McEwen~\textit{et al.}~\cite{mcewen2021resolving}, parameter regions of anomaly size, period, and frequencies of cosmic-ray strikes may change in the future. For example, they may be reduced by device-level improvement~\cite{martinis2021saving,pan2022engineering,iaia2022phonon}. 
While solid-state qubits on the substrates, such as silicon dots, color centers, and Majorana fermions, are expected to suffer from cosmic-ray strikes in a similar mechanism~\cite{mcewen2021resolving} and we can use the Q3DE for them, their parameter regions would be different from the superconducting qubits. 
According to the results in Fig.\,\ref{fig:scalability_estimate}, the Q3DE can reduce even if these parameters are improved in the future. Therefore, the Q3DE can reduce the hardware requirement in a variable region of parameters.

Trapped ions and neutral atoms are also promising candidates for scalable quantum devices. They use two stable states of atoms trapped in a vacuum by lasers as computational spaces. Since they do not reside on the substrate, the effect of cosmic rays would be negligible, but they are also expected to suffer from MBBEs caused by other factors, as shown in the next subsection.

\subsection{Other factors inducing MBBEs}
So far, we have focused on the MBBE events by cosmic rays because cosmic-ray-induced MBBEs are the dominant factor and are intensively studied. On the other hand, other promising devices, e.g., trapped ions and neutral atoms, are also expected to suffer from MBBEs induced by other factors.
In this section, we introduce three potential MBBEs of them and discuss how Q3DE can be applied to them.

First, trapped atoms may leak from traps due to fluctuations. The error probabilities of leaked atoms are effectively 50\% until they are reloaded. This event can be considered a single-bit burst error for neutral atoms. In the case of trapped ions, since they constitute a Coulomb crystal, all the ions in the crystal would be scrambled and become unavailable. Therefore, the effect for ions would be considered an MBBE event. In either case, our anomaly detection and decoder re-execution scheme would work for this case. Since we need actively reload atoms, we should move a logical qubit to another place instead of the code expansion. While the frequency of this event is once per two weeks in the state-of-the-art settings~\cite{dubielzig2021ultra}, stable controls sacrifice the control speeds and qubit integration. Therefore, the Q3DE enables an aggressive choice of parameters to speed up the computation as far as the MBBE frequencies are acceptable.
Note that if several burst errors occur on surface codes, the overhead of weighted matching may not be negligible. Nevertheless, it would not become a bottleneck since the lifetime of atoms is a few orders of magnitudes longer than superconducting qubits.
Second, atoms may transit to the stable states out of the qubit space. Once this event happens, the effective error rates of qubits become 50\% until they are re-pumped to the qubit space. According to the state-of-the-art controls of trapped ions~\cite{gaebler2016high}, the contribution of such leakage is about or below $10^{-5}$ per gate, which would be not negligible in the future since the current state-of-the-art applications require a few hours with $10^{3}$ logical qubits~\cite{babbush2018encoding,kivlichan2020improved,berry2019qubitization,lee2021even}. This error can also be considered a single-bit burst error and treated with the Q3DE.
The last MBBE is caused by a calibration drift of controls. In particular, the stray field of electrodes is reported as a non-negligible error in trapped ions~\cite{wineland1998experimental,liu2021minimization,ghadimi2022dynamic}. When the status of electrodes changes, the pre-calibrated trap controls cannot cool the atoms, which increases the error rates of trapped ions. This event can also be considered an MBBE. In this case, we need to move logical qubits to another space to perform the re-calibration.

In conclusion, we expect that the MBBE events exist in other candidates of qubits, and the Q3DE can be applied to them with small changes. For a more concrete estimation, further studies of parameter regimes with scaled devices is required, which we leave as future work.

\section{Conclusion}
\label{sec:conclusion}
In this paper, we proposed an FTQC architecture that is tolerant to MBBEs by cosmic rays, which are expected to be a severe barrier in the development of quantum computing. The proposed architecture can adaptively switch the QEC methods by three key processes: anomaly detection of MBBEs, dynamical code expansion, and re-execution of the error decoding process. We provided low-overhead algorithms and procedures for these components with theories of statistical modeling of the syndrome sequences and first-order analysis on the MBBE effect. The numerical results show that the proposed architecture can mitigate MBBEs with only a modest overhead. Thus, we believe that the Q3DE is a vital ingredient towards the attainment of a scalable FTQC and will open up the field of architecture design and compiler optimization for efficiently mitigating temporal error variations.

\section*{Acknowledgements}
This work was supported by JST PRESTO Grant Number JPMJPR1916 and JPMJPR2015, JST Moonshot R\&D Grant Number JPMJMS2061 and JPMJMS2067, JST ERATO Grant Number JPMJER1601, MEXT Q-LEAP Grant Number JPMXS0118068682, and JSPS KAKENHI Grant Number JP22H05000 and JP22K17868.

YS thanks Shuhei Tamate and Yutaka Tabuchi for the discussion on the superconducting qubits, Atsushi Noguchi on the ion traps, Keisuke Fujii on the qubit-allocation strategy, and Yosuke Ueno on the decoding architecture.

\newpage
\appendices
\section{Artifact Appendix}

\subsection{Abstract}
In this paper, we utilized several numerical simulations to evaluate the performance of the FTQC architecture with and without Q3DE. Also, we evaluated the performance of the decoding units with high-level synthesis, as shown in Table\,4. In this appendix, we explain how to regenerate the figures and tables used in the paper.

\subsection{Artifact check-list (meta-information)}

{\small
\begin{itemize}
  \item {\bf Program: }Numerical simulation codes, high-level description of circuits, and scripts to aggregate generated data and plot the figures.
  \item {\bf Compilation:} GCC 9.4.0
  \item {\bf Run-time environment: } Windows 11 for running Vitis HLS. Ubuntu 20.04 LTS on Windows Subsystem Linux for the other calculations.
  \item {\bf Output: }All the figures and parameters on tables in the main text.
  \item {\bf How much disk space required (approximately)?: }About 40 GB for intermediate output files. About 200 GB for the installation, mainly for Vitis HLS.
  \item {\bf How much time is needed to complete experiments (approximately)?: }About 6 days with 8 thread parallelization.
  \item {\bf Publicly available?: }Yes. All the source codes are uploaded to Zenodo.
  \item {\bf Code licenses (if publicly available)?: }Several codes are distributed with an original license. See \texttt{license.docx} in the published repository.
  \item {\bf Archived (provide DOI)?: } Yes. The assigned DOI is \texttt{10.5281/zenodo.7016156}.
\end{itemize}
}

\subsection{Description}
The codes are uploaded to Zenodo (DOI:~\texttt{10.5281/zenodo.7016156}).
We use a C++ compiler to build simulation codes. We verified that our code can be compiled with \texttt{g++ 9.4.0} on Windows Linux Subsystem and \texttt{Microsoft Visual Studio C++ 2019} on Windows 11.

For the environment-independent build, we use \texttt{CMake}.
The codes for generating figure 3,8 and figure 7 requires Edmonds' Blossom algorithms. We used Kolmogorov's implementation of the Blossom algorithm~\cite{kolmogorov2009blossom}. Since the license of this software does not allow public redistribution, please download the software from~\url{https://pub.ist.ac.at/~vnk/software.html}. Please see the workflow in the unzipped folder for how to allocate downloaded codes.
We use \texttt{Xilinx Vitis HLS} for the evaluation of the decoding unit. We used version 2021.2.1.

\subsection{Installation}
The programs are installed by downloading the zip file from the specified location and unzipping it. After that, we need to put the Blossom V to an appropriate directory, and need to install Vitis HLS with \texttt{Ultrascale+} devices.
See \texttt{README.md} in the published code repository.

\subsection{Experiment workflow}
The artifact consists of five folders, each of which can regenerate a figure or table. Figures 3 and 8, i.e., plots of logical errors, are generated from the same program with different configurations. 

All the experiments except for the high-level synthesis are performed by running a few shell and Python scripts. Commands are listed in the markdown document in each folder. The step-by-step workflow for regenerating parameters of high-level synthesis is also shown in the document in the corresponding folder.

\subsection{Evaluation and expected results}
As the result of the workflow, each folder generates a figure or parameters. These should be similar to the ones in the main text enough to result in the same conclusion.

Note that the figure would not be exactly the same since all the numerical simulation relies on the Monte-Carlo sampling for the evaluation, and there is a statistical fluctuation according to the number of sampling.

The regeneration of Table 4 requires several steps with \texttt{Vitis HLS 2021.2.1}. Please follow the instructions in the corresponding folder. The parameters of Table 4 also fluctuate according to the chosen random seed in HLS.

\newpage

\bibliographystyle{IEEEtranS}
\bibliography{refs}

\end{document}